\documentclass[twocolumn,aps,superscriptaddress,amsmath,amssymb]{revtex4-2}
\usepackage{amsfonts}
\usepackage{bbm}
\usepackage{graphicx}
\usepackage{epstopdf}
\usepackage{dcolumn}
\usepackage{bm}
\usepackage[colorlinks=true,linkcolor=blue,citecolor=blue]{hyperref}
\begin{document}
	
	\title{Strain-tunable electronic transport in disordered borophene nanoribbons with line defects}
	
	\author{Pei-Jia Hu}
	\affiliation{Hunan Key Laboratory for Super-microstructure and Ultrafast Process, School of Physics, Central South University, Changsha 410083, China}
	
	\author{Jia-Le Yu}
	\affiliation{Hunan Key Laboratory for Super-microstructure and Ultrafast Process, School of Physics, Central South University, Changsha 410083, China}
	
	\author{Tie-Feng Fang}
	\affiliation{School of Physical Science and Technology, Nantong University, Nantong 226019, China}
	
	\author{Ai-Min Guo}
	\email{aimin.guo@csu.edu.cn}
	\affiliation{Hunan Key Laboratory for Super-microstructure and Ultrafast Process, School of Physics, Central South University, Changsha 410083, China}
	
	\author{Qing-Feng Sun}
	\affiliation{International Center for Quantum Materials, School of Physics, Peking University, Beijing 100871, China}
	\affiliation{Hefei National Laboratory, Hefei 230088, China}
	
	\date{\today}
	
	\begin{abstract}
Resonant tunneling intrinsically exists in disordered borophene nanoribbons (BNRs) with randomly distributed line defects (LDs), providing a versatile platform for transport modulation.
Here, we theoretically investigate the effects of in-plane mechanical strain on the electronic transport properties of disordered BNRs coupled to metallic electrodes. Our results demonstrate that the resonant peaks in disordered BNRs of different widths remain robust against variations in electrode parameters and can be efficiently modulated via mechanical strain. Specifically, uniaxial strain significantly modulates the energies, widths, and amplitudes of these resonant peaks, and can even suppress the peaks selectively and induce new ones. In particular, the response of the resonant energies to the applied strain exhibits remarkable anisotropy, where strains along the longitudinal and transverse directions induce opposite energy shifts. Moreover, identical resonant peaks recurring in different nanoribbon widths respond uniformly to applied strain. Biaxial strain enables highly efficient conductance switching across multiple distinct levels, with the modulation frequency increasing for wider nanoribbons. Our findings establish a promising defect-strain-engineering strategy for manipulating electronic transport in borophene and other gapless materials.
	\end{abstract}
	
	\maketitle

	\section{Introduction}\label{sec1}

	The advent of two-dimensional (2D) materials, pioneered by the isolation of graphene, has revolutionized the landscape of nanoelectronics owing to their unique electronic structures and exceptional transport behaviors \cite{dassarma,gf2014,schwierz,roy}. Nevertheless, the lack of a bandgap in many 2D materials restricts effective transport modulation, despite their numerous advantages in other respects \cite{boland2024}. Over the past two decades, diverse strategies have been developed to tailor the electronic transport of 2D systems, which can be broadly categorized into intrinsic structural engineering and external physical modulation. Intrinsically, defect engineering, encompassing point defects such as atomic vacancies and extended line defects (LDs), can alter the local electronic environment and introduce novel scattering mechanisms \cite{jiang2019,liang2021,yu2024}. Externally, the application of physical controls, such as mechanical strain and electromagnetic fields, provides reversible knobs to modulate transmission channels \cite{dai2019,marchiori2022,qi2023,katiyar2024,li2024}. The synergistic combination of these intrinsic and external tuning strategies has given rise to a wealth of exotic electronic states and transport phenomena, thereby realizing multidimensional tailoring of quantum transport in 2D materials \cite{lehmann2013,zhang2022,zhang2023,wang2025,zhang2026}.
	
	Borophene, an atomically thin sheet of boron, has recently attracted extensive interest driven by its structural polymorphism, superior mechanical properties, and strong in-plane anisotropy \cite{mannix2018,kong2018,kaneti2021,ou2021,li2025}. Experimentally, numerous borophene phases, characterized by distinct hexagonal hole densities $\nu$, have been successfully synthesized on various metallic substrates \cite{ajm2015,bf2016,wl2018,rw2019,bk2019,nav2019,ch2020,rw2022,xl2022,zw2025,mx2026,sk2026}, including the notable $\nu_{1/5}$ and $\nu_{1/6}$ phases \cite{bf2016}. These two borophene phases exhibit mechanical flexibility and have been theoretically predicted to host exotic quantum phenomena, such as superconductivity and nontrivial topological states \cite{zhang2016,feng2017,ezawa2017,gao2017,shukla2018,gupta2018,wang2025second}. However, their inherent metallicity severely limits the tunability of electronic transport, hindering direct application in nanoelectronic devices.
	
\begin{figure}
		\centering
		\includegraphics[width=0.48\textwidth]{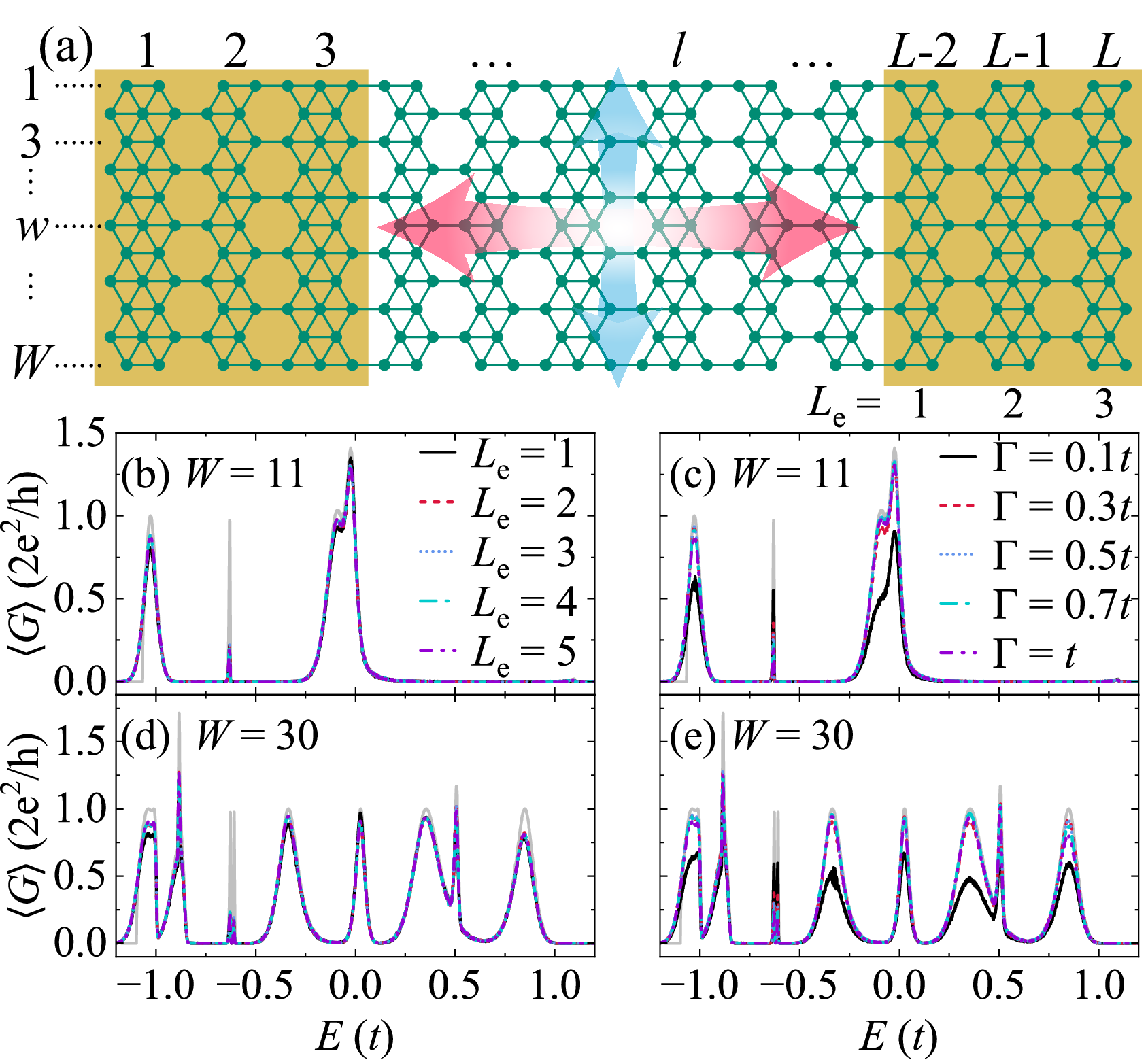}
		\caption{\label{fig1} Electronic transport in disordered BNRs with metallic electrode contacts. (a) Schematic of two-terminal disordered BNRs consisting of a random arrangement of (2,2) and (2,3) chains coupled to metallic electrodes at the yellow regions, where the red and blue double-headed arrows denote strain applied along the $x$ and $y$ directions, respectively. The length $L$ is defined by the total number of chains, the width $W$ is determined by the number of transverse atomic rows, and the contact length $L_e$ is the number of chains covered by each electrode. Here, $L=12$, $W=11$, and $L_e=3$.  Evolution of the transmission spectra with varying $L_e$ for (b) $W=11$ and (d) $W=30$ at a fixed contact strength of $\Gamma=t$.  Evolution of the transmission spectra with varying $\Gamma$ for (c) $W=11$ and (e) $W=30$ at $L_e=3$. The gray lines in (b)-(e) represent the results calculated by replacing the wide-band metallic electrodes with semi-infinite $\nu_{1/5}$ BNR leads.}
	\end{figure}

	An unconventional LD featuring perfect lattice matching has been widely observed to naturally emerge at the interface between the $\nu_{1/5}$ and $\nu_{1/6}$ phases \cite{liu2018,wang2020,kong2019,li2021,ql2024}. The structural units of these phases, namely the (2,2) and (2,3) chains, serve as fundamental building blocks to construct novel mixed-phase borophenes, where $(n_n, n_w)$ denotes the number of atoms in the narrow and wide rows of a single boron chain, respectively [Fig.~\ref{fig1}(a)]. Specifically, Xu \textit{et al.} demonstrated that LDs unexpectedly enhance both the structural stability and mechanical response of borophene, driving the system to energetically favor long-range disorder \cite{xu2021}. Yu \textit{et al.} revealed that the formation of LDs in borophene is driven by a critical synergy between the in-plane migration of boron atoms and their sinking into the substrate, rationalizing the experimental temperature window for its synthesis \cite{yu2026}. Our previous studies demonstrated that, despite the overall suppression of electronic transport in disordered borophene nanoribbons (BNRs) with randomly distributed LDs, robust quantized resonant peaks persist regardless of the nanoribbon length, the LD ratio, and their arbitrary arrangement \cite{s41524,hu2023}. Notably, these findings provide a theoretical explanation for the experimental observation of one-dimensional nearly free electron states induced by LDs \cite{kong2019}.
In stark contrast to the intrinsic metallicity of pristine borophene, which severely limits its transport tunability, the resonant tunneling on an insulating background in disordered BNRs lays a solid foundation for subsequent transport modulation \cite{prb109}.

	In this paper, we theoretically investigate the effects of in-plane mechanical strain on the electronic transport properties of two-terminal disordered BNRs coupled to metallic electrodes. Using the tight-binding model combined with the Green's function method, we demonstrate that the resonant peaks in disordered BNRs remain robust against variations in electrode parameters and can be effectively tuned by applied strain. Specifically, uniaxial strain significantly modulates the energies, widths, and amplitudes of the resonant peaks, and can even selectively suppress existing peaks and induce new ones. Furthermore, the response of the resonant energies exhibits a striking anisotropy, with longitudinal and transverse strains inducing opposite energy shifts. By analyzing the local current distributions, we reveal the preservation of spatial inversion symmetry under strain, which ensures that recurrent resonant peaks across different nanoribbon widths share a uniform mechanical response. Finally, we show that highly efficient transport modulation is achieved under biaxial strain with equal magnitudes but opposite signs, enabling the conductance to switch between multiple distinct levels, with the modulation frequency increasing in wider nanoribbons. Overall, by synergizing intrinsic LDs with external mechanical strain, our study provides an effective strategy for tailoring electron transport in gapless low-dimensional materials, advancing their potential for next-generation nanoelectronics.
	
	The remainder of this paper is organized as follows. In Sec.~\ref{sec2}, we introduce the model Hamiltonian and the theoretical formalism for electronic transport. In Sec.~\ref{sec3}, we investigate the effects of metallic contacts, as well as uniaxial and biaxial strains, on electronic transport in disordered BNRs, and explore the potential for device applications. Finally, we summarize our findings in Sec.~\ref{sec4}.
	
	\section{Model and Method}\label{sec2}

	As illustrated in Fig.~\ref{fig1}(a), the disordered BNR, constructed through a random arrangement of (2,2) and (2,3) atomic chains, is described by the tight-binding Hamiltonian
	\begin{equation}
		\mathcal{H} = \sum_{i}\epsilon_{i}a_{i}^{\dagger}a_{i} - \sum_{\langle ij\rangle}t_{ij}a_{i}^{\dagger}a_{j},
	\end{equation}
where $a_{i}^{\dagger}$ ($a_{i}$) is the creation (annihilation) operator for an electron at site $i$, $\epsilon_{i}$ represents the on-site energy, and $t_{ij}$ denotes the nearest-neighbor hopping integral between adjacent sites $i$ and $j$. Given the comparable mechanical properties of the constituent $\nu_{1/5}$ and $\nu_{1/6}$ phases, it is a reasonable approximation to treat the structural deformation of the mixed-phase BNR as uniform under moderate in-plane strains \cite{shukla2018,xu2021}.
The atomic coordinates of the nanoribbon are rescaled as $x_i' = x_i(1 + s_x)$
and $y_i' = y_i(1 + s_y)$, where ($x_i$, $y_i$) are the coordinates of atom $i$, and $s_x$ and $s_y$ (collectively denoted as $\boldsymbol{s} = (s_x, s_y)$) represent the strain components along the
$x$ and $y$ directions, respectively [see the red and blue double-headed arrows in Fig.~\ref{fig1}(a)].
To capture this deformation, we adopt a distance-dependent hopping integral $t_{ij} = t_{ij}^0 \exp\left[-\beta \left(d_{ij}/d_{0} - 1\right)\right]$, where $t_{ij}^0$ is the unstrained hopping integral, $d_{ij}$ and $d_{0}$ represent the strained and unstrained interatomic bond lengths, respectively, and $\beta$ denotes the electron Gr\"{u}neisen parameter \cite{davoudiniya2021}.
	
	Using the Landauer-Büttiker formula combined with the Green’s function method, the two-terminal conductance is given by \cite{sd1997}
	\begin{equation}\label{eq2}
		G = \frac{2e^2}{h} \operatorname{Tr} \left( \boldsymbol{\Gamma}_{\rm S} \mathbf{G}^{\rm r} \boldsymbol{\Gamma}_{\rm D} \mathbf{G}^{\rm a} \right),
	\end{equation}
	where $\mathbf{G}^{\rm r} = \left( E\mathbf{I} - \mathbf{H} - \boldsymbol{\Sigma}_{\rm S}^{\rm r} - \boldsymbol{\Sigma}_{\rm D}^{\rm r} \right)^{-1}$ is the retarded Green's function with $\boldsymbol{\Sigma}_{\rm S/D}$ the self-energies, and $\boldsymbol{\Gamma}_{\rm S/D} = i \left( \boldsymbol{\Sigma}_{\rm S/D}^{\rm r} - \boldsymbol{\Sigma}_{\rm S/D}^{\rm a} \right)$ are the linewidth functions describing the coupling to the source/drain electrodes. To emulate a realistic contact architecture tailored for practical device applications, we consider metallic wide-band electrodes with an energy-independent coupling strength $\Gamma$ and a contact length $L_e$ [Fig.~\ref{fig1}(a)]. The bond current between adjacent sites $i$ and $j$ is given by
	\begin{equation}\label{eq3}
		I_{ij} = -\frac{2e}{h} \int_{-\infty}^{\infty} \text{Re} \left[ \mathbf{H}_{ij} \mathbf{G}^{<}_{ji} \right] dE,
	\end{equation}
	where $\mathbf{G}^{<}$ is the lesser Green's function. Numerically, $\mathbf{G}^{\rm r}$ and $\mathbf{G}^{<}$ are efficiently calculated using the recursive Green's function technique \cite{chl}.

	\section{Results and Discussion}
	\label{sec3}
	For transport calculations, we implement inhomogeneous tight-binding parameters with $\epsilon_i$ and $t_{ij}^0$ determined by the local coordination numbers \cite{prb109}. The Gr\"{u}neisen parameter is set to $\beta = 3$, and our main conclusions remain valid for $\beta$ within a physically reasonable range \cite{pereira2009,pearce2016,quereda2016,fang2018}. The conductance is calculated for disordered BNRs of length $L = 1000$ composed of randomly distributed (2,2) and (2,3) chains with equal probabilities, and averaged over 2500 independent disorder configurations.
	
	Given that borophene is experimentally synthesized on metallic substrates and inevitably integrated with metal contacts in realistic devices, it is essential to account for the contact effects of metal electrodes \cite{scopel2023}. Figures~\ref{fig1}(b) and \ref{fig1}(d) show the averaged conductance $\langle G \rangle$ as a function of energy $E$ for $W=11$ and $W=30$, respectively, under varying contact lengths $L_e$ at a fixed
electrode-borophene coupling strength of $\Gamma=t$. 
Although the overall conductance is severely suppressed by the strong interface scattering induced by the random arrangement of LDs, several conductance peaks remain well preserved \cite{s41524,prb109}. In contrast to the semi-infinite $\nu_{1/5}$ BNR leads, which yield perfectly quantized peaks with a conductance of $G_0 = 2e^2/h$ [see the gray lines in Figs.~\ref{fig1}(b) and \ref{fig1}(d)], coupling to metallic electrodes causes the resonant peaks to undergo varying degrees of suppression depending on the peak widths [see the other lines in Figs.~\ref{fig1}(b) and \ref{fig1}(d)]. This phenomenon can be attributed to interface scattering induced by the mismatch between the electrode and resonance modes, which, under the wide-band approximation, is primarily determined by the spatial profiles of the specific resonant states.
	
	For $W=11$, four resonant peaks emerge within the considered energy window $E \in [-1.2t, 1.2t]$, including a broad peak at $E = -1.028t$, a sharp peak at $E = -0.630t$, and two adjacent peaks at $E = -0.095t$ and $-0.024t$ [Fig.~\ref{fig1}(b)]. In the presence of metallic electrodes, the sharp peak is significantly suppressed, with its conductance dropping to $0.23G_0$, whereas the wider resonant peaks are well preserved. This peak-width-dependent effect becomes even more pronounced for $W=30$, where four sharp peaks are strongly suppressed compared to the others [Fig.~\ref{fig1}(d)]. However, the conductance of all resonant peaks remains nearly unchanged as $L_e$ increases from 1 to 5, suggesting strong robustness against contact area variations in experimental device fabrication.

	To emulate the contacts of different metal interfaces, Figs.~\ref{fig1}(c) and \ref{fig1}(e) display the transmission spectra for $W=11$ and $W=30$, respectively, under varying coupling strengths $\Gamma$ at $L_e=3$. Under a weak contact strength of $\Gamma=0.1t$, all resonant peaks are suppressed, with the conductance of isolated ones dropping to approximately $0.6G_0$ [see the black lines in Figs.~\ref{fig1}(c) and \ref{fig1}(e)]. As $\Gamma$ increases, the sharper peaks are further suppressed, whereas the wider ones are enhanced [see the other lines in Figs.~\ref{fig1}(c) and \ref{fig1}(e)]. This behavior can be attributed to unsaturated electron injection at weak contact strengths, whereas interface scattering dominates at stronger couplings. Overall, our calculations demonstrate that the resonant tunneling in disordered BNRs is experimentally observable in practical device measurements, remaining robust against variations in electrode parameters. Without loss of generality, we fix $L_e=3$ and $\Gamma=t$ in the subsequent calculations.

	\begin{figure}
		\centering
		\includegraphics[width=0.48\textwidth]{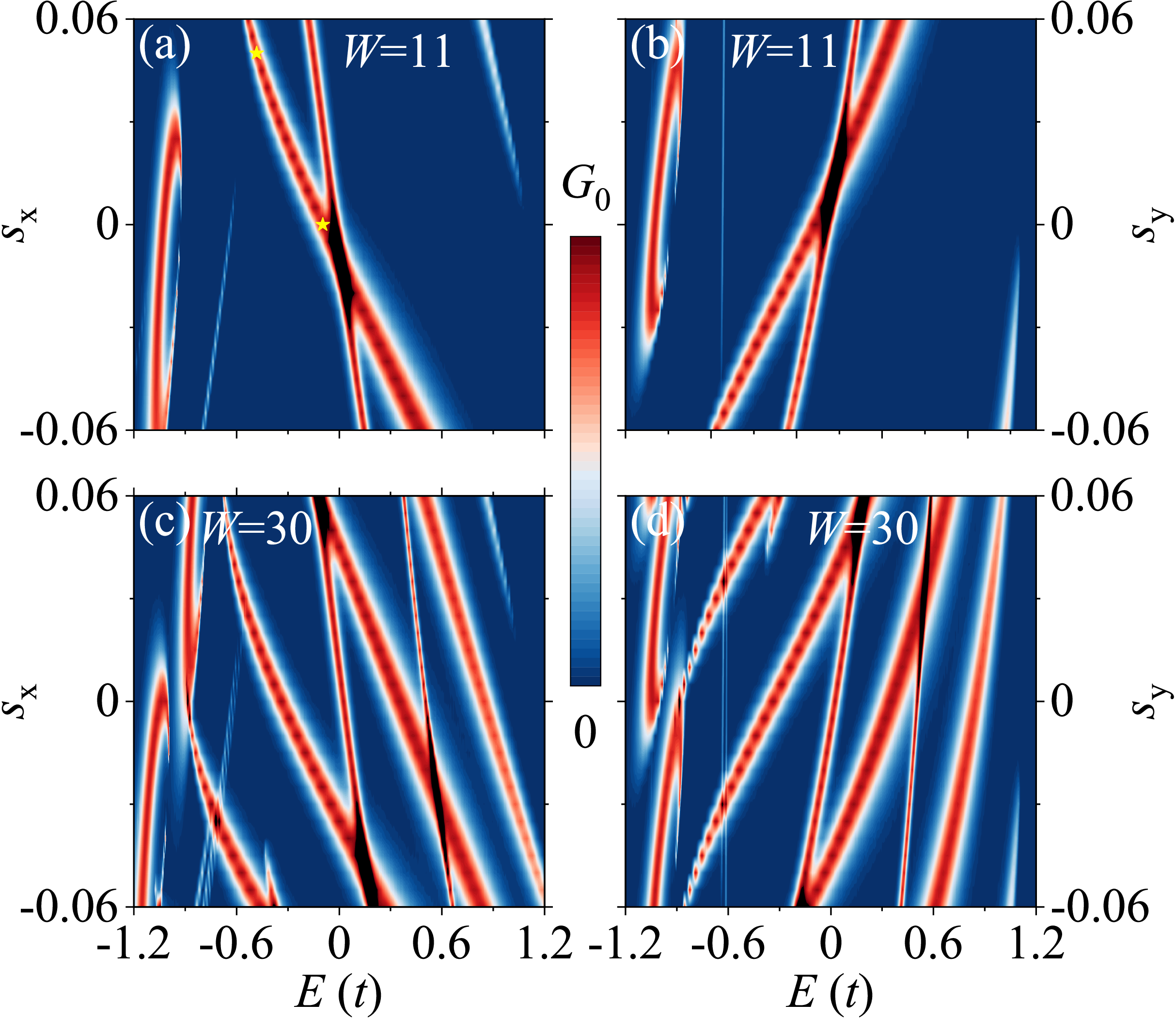}
		\caption{\label{fig2} Electronic transport in disordered BNRs under uniaxial strain. Evolution of the averaged conductance $\langle G \rangle$ as functions of energy $E$ and strain for disordered BNRs with widths of (a), (b) $W=11$ and (c), (d) $W=30$. The strain is applied along the $x$-direction $s_x$ in panels (a) and (c), and along the $y$-direction $s_y$ in panels (b) and (d).}
	\end{figure}
	
	To systematically investigate the strain effects on electronic transport, Figs.~\ref{fig2}(a) and \ref{fig2}(b) present the evolution of transmission spectra as a function of uniaxial strain $s_x$ and $s_y$, respectively, for $W=11$. The black, red, and white trajectories on the blue background track the strain-dependent evolution of the resonant peaks, with their thickness directly reflecting the peak widths. It is clear that the energies of the resonant peaks are highly sensitive to the applied strain, exhibiting a strong anisotropic response. For instance, the two adjacent peaks initially located at $E = -0.095t$ and $-0.024t$ without strain shift to lower (higher) energies with increasing (decreasing) $s_x$ [Fig.~\ref{fig2}(a)], whereas the opposite trend is observed for $s_y$ [Fig.~\ref{fig2}(b)]. Accompanying these energy shifts, the peak widths also evolve correspondingly. Notably, resonant peaks with larger widths exhibit greater robustness against strain than sharper ones, remaining stable across a broad range of applied strain. Moreover, these resonant peaks exhibit different sensitivities to both strain $s_x$ and $s_y$, with the peak at $E = -0.095t$ exhibiting a more pronounced energy shift than that at $-0.024t$. Consequently, these two peaks intersect at $s_x = -0.015$ or $s_y = 0.015$ and diverge as the strain deviates from these values, forming a distinct X-shaped trajectory. Beyond the shift of resonant energies, the applied strain can also suppress existing peaks selectively and induce new ones. For example, with increasing $s_x$ or decreasing $s_y$, the resonant peaks near $E = -t$ vanish while a new peak emerges near $E = t$.
	
	Figures~\ref{fig2}(c) and \ref{fig2}(d) present the evolution of transmission spectra for $W=30$ under uniaxial strains $s_x$ and $s_y$, respectively. Consistent with the $W=11$ case, the applied strain leads to prominent energy shifts of all resonant peaks, accompanied by the selective suppression of existing peaks and the emergence of new ones. Owing to the increased peak density and the different energy shift sensitivities of individual peaks, a complex mesh-like evolution pattern emerges. Despite this visual complexity, the fundamental anisotropic response remains robust, with the peaks consistently shifting in opposite directions under $s_x$ versus $s_y$. These results reveal that resonant tunneling in disordered BNRs can be precisely tuned via mechanical strain, highlighting their potential for applications in strain-gated electronic switches.
	
	\begin{figure}
		\centering
		\includegraphics[width=0.48\textwidth]{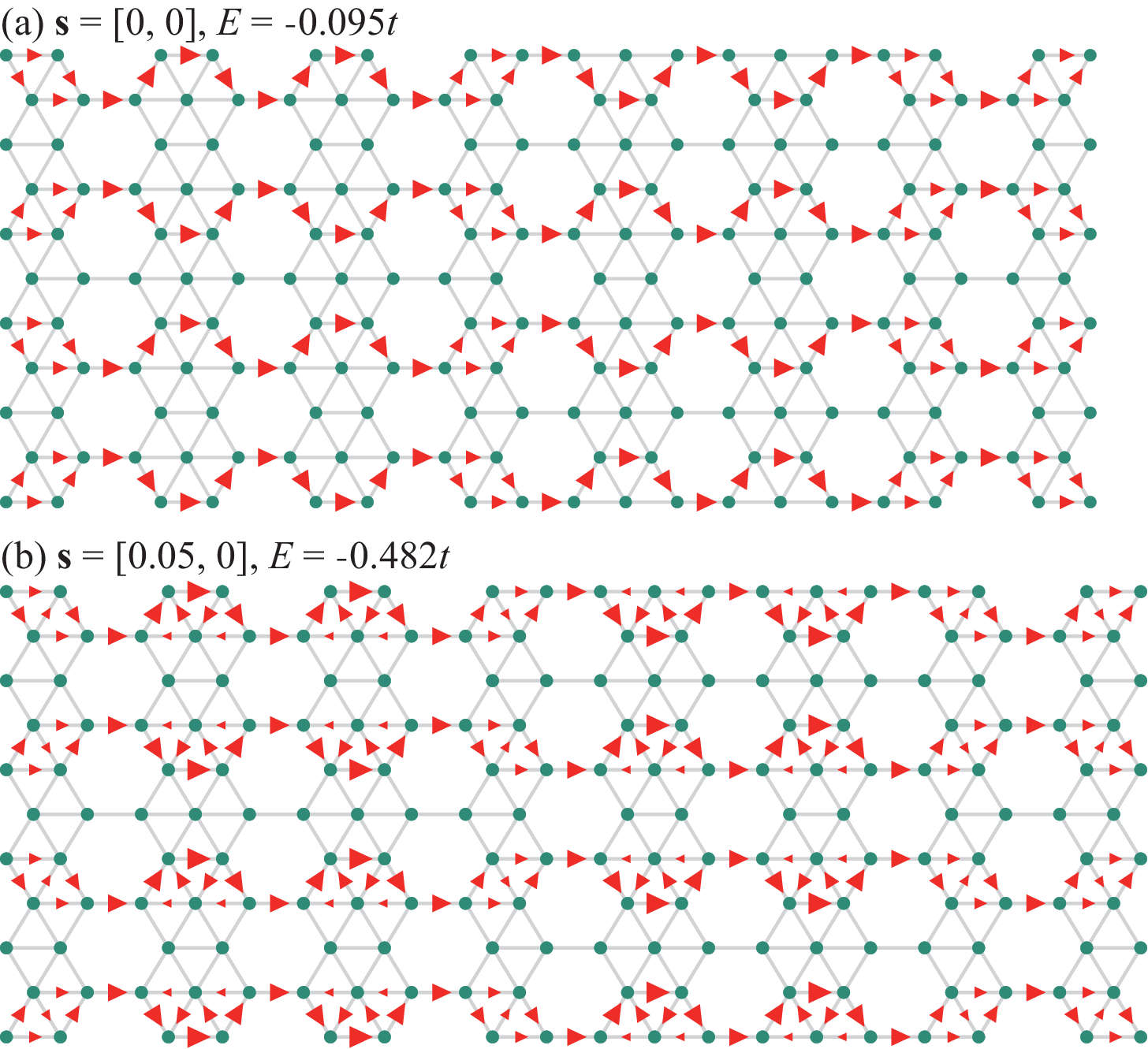}
		\caption{\label{fig3} Local current density distributions for (a) $\boldsymbol{s}= (0, 0)$ at $E = -0.095t$ and (b)  $\boldsymbol{s}= (0.05, 0)$ at $E = -0.482t$, corresponding to the resonant peaks marked by yellow stars in Fig.~\ref{fig2}(a). The distributions are shown for a short segment containing eight chains of both (2,2) and (2,3) types, extracted from a disordered configuration with $L = 1000$. Red arrows denote the bond currents, with their orientations and sizes reflecting the current direction and magnitude, respectively.}
	\end{figure}
	
	Figures~\ref{fig3}(a) and \ref{fig3}(b) display the local current density distributions at $\boldsymbol{s}= (0, 0)$, $E = -0.095t$ and $\boldsymbol{s}= (0.05, 0)$, $E = -0.482t$, respectively. Here, the peak at $E = -0.482t$ evolves from that at $E = -0.095t$, corresponding to the yellow stars in Fig.~\ref{fig2}(a). For $E = -0.095t$, the current flows along four narrow channels of width 2, exhibiting spatial inversion symmetry with respect to the center of the nanoribbon [Fig.~\ref{fig3}(a)]. Within each narrow channel, the current always flows along the rhombic edges of the (2,2) chains and the armchair edges of the (2,3) chains without undergoing scattering at the interfaces, demonstrating resonant tunneling. For $E = -0.482t$, the current remains confined to four narrow channels of width 2, dictated by the preservation of spatial inversion symmetry in the disordered BNR during stretching [Fig.~\ref{fig3}(b)]. However, the current within each chain redistributes, flowing not only along the edges but also across the inner lattice connections. This result signifies that the propagation mode changes during stretching, thereby elucidating the variation in resonant peak widths, as well as the selective suppression of existing peaks and the induction of new ones.
	
	\begin{figure*}[htbp]
		\centering
		\includegraphics[width=1\textwidth]{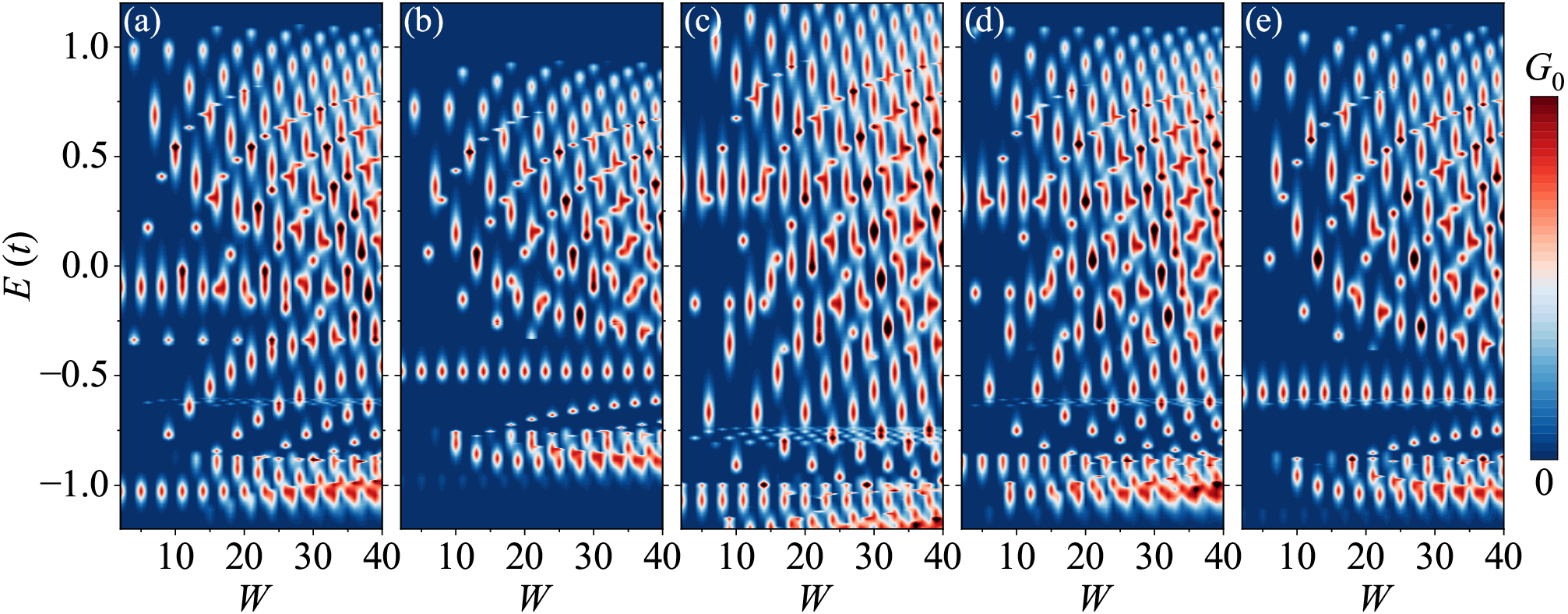}
		\caption{\label{fig4} Width-dependent electronic transport in disordered BNRs under various uniaxial strains. $\langle G \rangle$ as functions of energy $E$ and nanoribbon width $W$ under different applied strains, with (a) $\boldsymbol{s} = (0, 0)$, (b) $\boldsymbol{s} = (0.05, 0)$, (c) $\boldsymbol{s} = (-0.05, 0)$, (d) $\boldsymbol{s} = (0, 0.05)$, and (e) $\boldsymbol{s} = (0, -0.05)$.}
	\end{figure*}
	
	To demonstrate the universality of the strain-induced modulation of the resonant peaks, Figs.~\ref{fig4}(a)-\ref{fig4}(e) display the evolution of the transmission spectra as a function of the nanoribbon width $W$ under different strains of $\boldsymbol{s} = (0, 0)$, $\boldsymbol{s} = (0.05, 0)$, $\boldsymbol{s} = (-0.05, 0)$,  $\boldsymbol{s} = (0, 0.05)$, and $\boldsymbol{s} = (0, -0.05)$, respectively. In the absence of strain, the resonant peaks are present across nanoribbons of varying widths, with the peak density increasing as the nanoribbon widens [Fig.~\ref{fig4}(a)]. Notably, the identical resonant peak, characterized by the same resonant energy and peak width, is observed across different nanoribbon widths, obeying a specific recurrence rule $W=(W_i + 1)n - 1$ \cite{s41524,prb109}. Here, $W_i$ denotes the initial width at which a specific resonant peak emerges, and $n$ is a positive integer. For instance, the resonant peak at $E=-0.095t$ first appears at $W=2$ and periodically recurs at $W=3n-1$. Tracing back to the local current distributions in Fig.~\ref{fig3}(a), the microscopic physical meaning of $W_i$ essentially corresponds to the width of narrow current channels.
	
	Upon the application of strain, the resonant energies undergo noticeable and anisotropic shifts, while the population of resonant peaks remains nearly unchanged. Overall, the global energy window of the resonant peaks is highly sensitive to strain along the $x$-direction, compressing under $\boldsymbol{s} = (0.05, 0)$ and broadening under $\boldsymbol{s} = (-0.05, 0)$ [Figs.~\ref{fig4}(b) and \ref{fig4}(c)], but remains largely unaffected by strain applied along the $y$-direction [Figs.~\ref{fig4}(d) and \ref{fig4}(e)]. The recurrence rule $W=(W_i + 1)n - 1$ for a peak initially emerging at $W_i$ is strictly preserved under applied strain, due to the preservation of spatial inversion symmetry [Fig.~\ref{fig3}(b)]. As an example, the resonant peaks satisfying $W = 3n - 1$ emerge at $E = -0.482t$ for $\boldsymbol{s} = (0.05, 0)$ [Fig.~\ref{fig4}(b)], at $E = 0.376t$ for $\boldsymbol{s} = (-0.05, 0)$ [Fig.~\ref{fig4}(c)], at $E = 0.312t$ for $\boldsymbol{s} = (0, 0.05)$ [Fig.~\ref{fig4}(d)], and at $E = -0.579t$ for $\boldsymbol{s} = (0, -0.05)$ [Fig.~\ref{fig4}(e)], all of which evolve from the resonant peaks at $E=-0.095t$ for $\boldsymbol{s} = (0, 0)$. These results reveal a width-independent strain response for recurring resonant peaks, enabling the understanding of strain-tunable electronic transport behaviors from narrow to extremely wide nanoribbons.

	\begin{figure}
		\centering
		\includegraphics[width=0.48\textwidth]{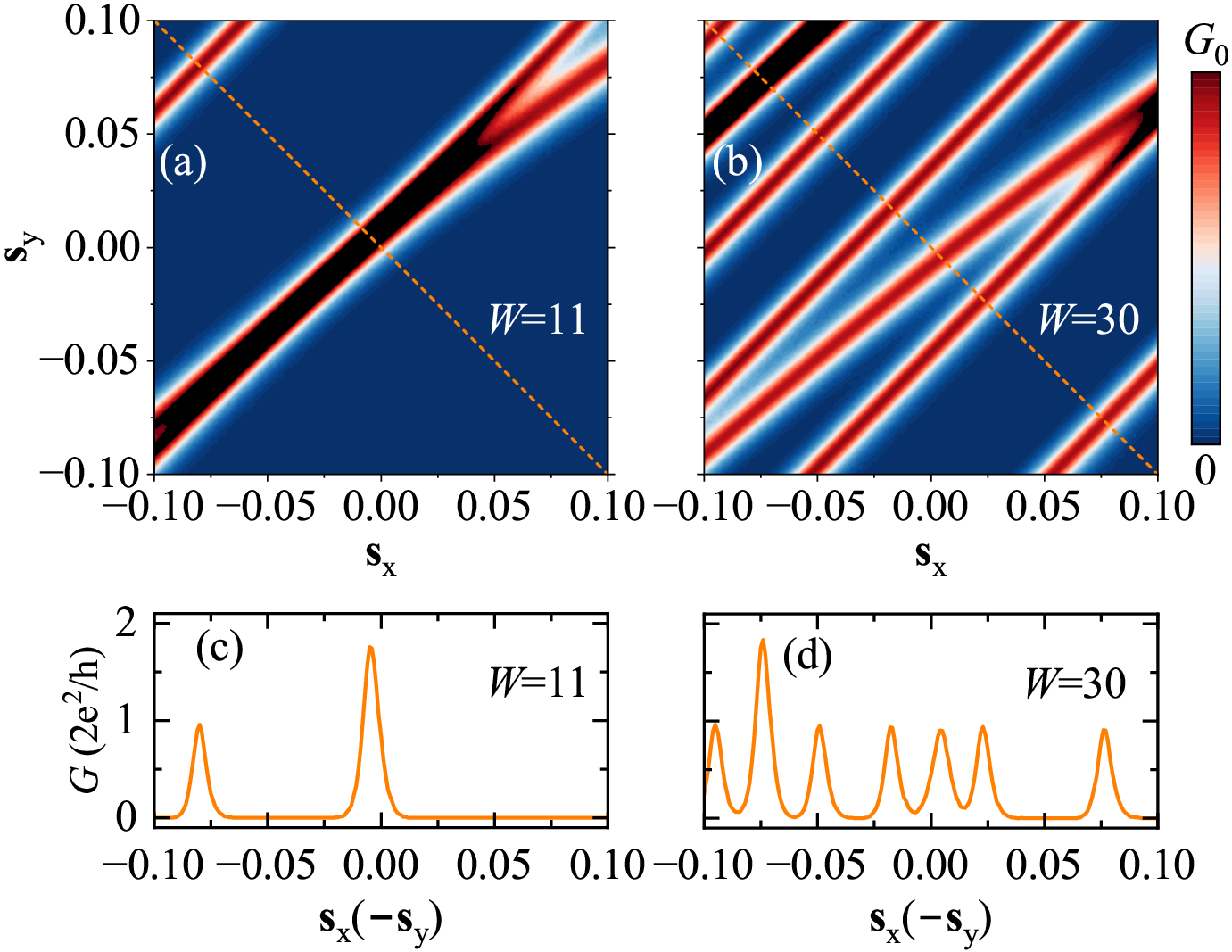}
		\caption{\label{fig5} Electronic transport in disordered BNRs under biaxial strain.  $\langle G \rangle$ at $E=0$ as functions of $s_x$ and $s_y$ for nanoribbon widths of (a) $W=11$ and (b) $W=30$. $\langle G \rangle$ at $E=0$ as functions of $s_x$ under the constraint $s_y=-s_x$ for (c) $W=11$ and (d) $W=30$, respectively, corresponding to the paths marked by the orange dashed lines in (a) and (b).}
	\end{figure}
	
	Finally, we investigate the electronic transport in disordered BNRs subjected to biaxial strain. Figures~\ref{fig5}(a) and \ref{fig5}(b) present 2D plots of $\langle G \rangle$ at $E=0$ as functions of $s_x$ and $s_y$ for $W=11$ and $W=30$, respectively. While $E=0$ is taken as a representative case, similar phenomena hold for other energies. For $W=11$, two linear regions with high conductance emerge for strains satisfying $s_y = s_x + \delta$ [Fig.~\ref{fig5}(a)]. This is because the strains along the $x$- and $y$-directions have opposite modulation effects on the resonant energies, and thus they effectively cancel each other out when $s_y = s_x$, keeping the resonant energies nearly unchanged. For $s_y = s_x + \delta$, the resonant energies shift to $E=0$, giving rise to the high-conductance linear regions. Therefore, $\delta$ is proportional to the energy difference between the unstrained resonant peaks and $E=0$. A similar phenomenon is observed for $W=30$, albeit with a greater number of high-conductance linear regions emerging due to the increased density of resonant peaks [Fig.~\ref{fig5}(b)].
	
	To achieve highly efficient transport modulation, it is natural to consider biaxial strain applied in opposite directions. Figures~\ref{fig5}(c) and \ref{fig5}(d) show the evolution of $\langle G \rangle$ at $E=0$ for $W=11$ and $W=30$ along the strain path $s_y = -s_x$. For $W=11$, the conductance can be tuned across three distinct levels, a high level of $1.76 G_0$, an intermediate level of $0.96 G_0$, and $0$ [Fig.~\ref{fig5}(c)]. Beyond serving as a conventional switch with a high on-off ratio, this strain-tunable three-state behavior serves as a fundamental building block for multivalued logic \cite{andreev2022}. For $W=30$, the conductance can similarly switch among distinct levels under biaxial strain [Fig.~\ref{fig5}(d)]. More importantly, due to the increased density of resonant peaks, the modulation frequency is significantly enhanced. Owing to the periodic recurrence of resonant peaks across varying widths, disordered BNRs provide a versatile platform for designing strain-tunable switches with tailored modulation frequencies.
	
	\section{Conclusion} \label{sec4}	
	In summary, we theoretically demonstrate that resonant peaks in disordered BNRs of varying widths remain robust against metal electrode parameters and can be efficiently tuned via mechanical strain. Specifically, uniaxial strain drives highly anisotropic shifts in the resonant energies, accompanied by the selective suppression of existing peaks and the emergence of new ones. Analysis of local current distributions reveals that this strain-induced modulation is highly consistent across varying nanoribbon widths, governed by the preservation of spatial inversion symmetry and obeying a strict recurrence rule. Furthermore, biaxial strain maximizes this modulation, enabling the conductance to switch between multiple distinct levels, with the modulation frequency significantly increasing in wider nanoribbons. The defect-strain-engineering strategy establishes disordered BNRs as a compelling physical platform for multivalued logic architectures, offering a paradigm for other gapless materials.
	
	\section*{Acknowledgments}
	This work is supported by the National Natural Science Foundation of China (Grants No. 12674067, No. 12274466, No. 12374034, and No. 12474045), the
	Quantum Science and Technology-National Science and Technology Major Project (Grant No. 2021ZD0302403), the National Key R and D Program of China (Grant No. 2024YFA1409002), and the High Performance Computing Center of Central South University.

\end{document}